\documentclass[aps,prstab,groupedaddress,showpacs,showkeys,twocolumn,floatfix]{revtex4-1}

\usepackage{graphicx} \usepackage{url} \usepackage{amsmath}
\usepackage{color}
\usepackage{psfrag}

\begin{document}

\title{Simultaneous  linear optics and coupling correction for storage rings with 
turn-by-turn beam position monitor data}

\author{Xi Yang }
\affiliation{Brookhaven National Laboratory, Upton, Long Island, NY 11973, USA}

\author{Xiaobiao Huang }
\email[]{xiahuang@slac.stanford.edu} 
\affiliation{SLAC National Accelerator Laboratory, Menlo Park, CA 94025}

\date{\today}

\begin{abstract}
We propose a method to simultaneously correct linear optics errors and linear coupling 
for storage rings using turn-by-turn (TbT) beam position monitor (BPM) data. 
The independent component analysis (ICA) method is used to isolate the betatron normal modes 
from the measured TbT BPM data. The betatron amplitudes and  phase advances of the 
projections of the normal modes on the horizontal and vertical planes are then extracted, 
which, combined with dispersion measurement, are used to fit the lattice model. The fitting 
results are used for lattice correction. 
The method has been successfully demonstrated on the NSLS-II storage ring. 

\end{abstract}
 

\maketitle

\section{\label{intro}Introduction}
Linear optics correction has crucial importance in the operation of a storage ring accelerator. 
There are many error sources that contribute to deviations of the storage ring optics 
from the ideal model. 
These include systematic and random errors of quadrupole components of all magnets, feeddown from horizontal 
orbit offsets in sextupole magnets, and perturbations due to insertion devices. 
Linear optics errors usually degrade the nonlinear dynamics performance of the storage ring, causing a reduction 
of dynamic aperture and momentum aperture. 
Linear optics errors can be corrected with adjustments to the strengths of quadrupole magnets. 
Correction of linear optics often lead to improvements of injection efficiency and/or Touschek lifetime~\cite{SafranekLOCO}.  
It may also be necessary to correct the linear optics in order to deliver certain beam parameters 
to facilitate user experiments, beam diagnostics, or machine protection. 
For example, accurate beta functions may be required at certain locations of the ring, or an accurate phase 
advance may be required between two storage ring components. 

Linear coupling between the horizontal and vertical planes and spurious vertical dispersion are 
other types of common errors in 
a storage ring that need to be controlled.
Linear coupling can be caused by skew quadrupole components of magnets through magnet errors, rolls of 
quadrupoles, and vertical orbit offset in sextupole magnets. Spurious vertical dispersion can 
be caused by vertical steering magnets and coupling of the horizontal dispersion through skew quadrupole components 
in dispersive regions. 
Both linear coupling and spurious vertical dispersion contribute to the vertical emittance and 
both can be corrected with skew quadrupoles. The reduction of vertical emittance through linear coupling and spurious vertical 
dispersion correction is often referred to as ``coupling correction''. 

Linear optics and coupling correction for storage rings is typically done with the orbit response matrix 
based method, LOCO (linear optics from closed orbit)~\cite{SafranekLOCO}.  
By fitting quadrupole and skew quadrupole variables in the lattice model to the measured orbit response matrix 
and dispersion data, 
LOCO finds a set of magnet errors that can give rise to the observed lattice errors. 
Correcting the magnet errors in the machine then leads to improved linear optics and reduced coupling error. 

In recent years turn-by-turn (TbT) BPMs have become widely used in storage rings. 
TbT BPMs not only detect the closed orbit, but also the orbit of a beam in coherent oscillation. 
From the latter betatron amplitudes and phase advances can be derived~\cite{CastroPAC93, WangCX2003, XHuangICA2005}, 
which in turn can be used for optics correction~\cite{XHuangICA2005,AibaLHC.2009.PhysRevSTAB.12.081002,ShenRHIC2013}.
TbT BPM data also contain linear coupling information and can be used for coupling correction. 
A method based on the correction of the global linear coupling resonance driving term and leading Fourier harmonics 
of vertical dispersion function was proposed in Ref.~\cite{WangF.PhysRevSTAB.11.050701}. 
A method based on the correction of coupling resonance driving terms was proposed and demonstrated in 
Ref.~\cite{FranchiCoupling}.  
Both methods require linear optics correction beforehand in order to obtain an accurate optics model as needed 
for coupling correction.
Ref.~\cite{HuangFitTBT2010} proposed a method that could be used to simultaneously correct linear optics and coupling. 
One disadvantage of this method is that the BPM errors of the first two BPMs propagate downstream and may affect 
the fitting results. 

In this paper we propose and experimentally demonstrate a new method to simultaneously correct 
linear optics and coupling. 
The independent component analysis (ICA) method is first applied to extract the 
amplitudes and phases of the projection of the normal modes 
on the horizontal and vertical BPMs~\cite{XHuangICA2005}, 
which are then compared to their model generated counterparts in fitting. 
The fitting scheme is similar to LOCO. 
Since closed orbit response and coherent orbit oscillation sample the optics and coupling errors of the machine 
in a similar fashion, it is expected the performance of this method would be similar to that of LOCO. 
However, the TbT BPM data based method has a great advantage in that data taking is significantly faster than 
LOCO. The time for taking orbit response matrix data may vary from ~10-100 minutes for different machines, 
while TbT BPM data taking takes only a few seconds. 
Simulation results for our new method were previously reported in Ref.~\cite{HuangYangIPAC15}. 

In the following we first describe the method in section~\ref{secMethod}.  
A discussion of simulation results is in section~\ref{secMethod}. 
Experimental results on the National Synchrotron Light Source-II (NSLS-II) storage ring 
are presented in section~\ref{secExperi}. Conclusion is given in section~\ref{secConclu}. 

\section{Optics and coupling correction with ICA \label{secMethod}} 
Betatron motion with linear coupling can be decoupled into two normal modes~\cite{EdwardsTeng,SaganRubin}.
In general, the beam motion observed by a BPM on any of the two transverse planes has components of both normal modes. 
Normally the two modes have different betatron tunes and hence can be separated with 
the ICA method when TbT BPM data from BPMs around the ring are analyzed 
together~\cite{XHuangICA2005}.   
In the ICA process, BPM noise is reduced and other components of beam motion, such as synchrotron motion and nonlinear 
resonance terms, are isolated from the betatron motion. 
Therefore, the resulting betatron components have high accuracy. 

Each betatron normal mode corresponds to two orthogonal ICA modes. The betatron components 
on each BPM consist of four ICA modes, which can be expressed as 
\begin{eqnarray}
    x_n &=& A \cos\Psi_{1n} -  B \sin\Psi_{1n} +c \cos\Psi_{2n} -  d \sin\Psi_{2n}, \nonumber \\
    y_n &=& a \cos\Psi_{1n} -  b \sin\Psi_{1n} +C \cos\Psi_{2n} -  D \sin\Psi_{2n},   \label{eq:ICAmodes}
\end{eqnarray}
where $x_n$ and $y_n$ are observed beam positions on the horizontal and vertical planes at 
the $n$'th turn, respectively, 
$\Psi_{1n,2n}=2\pi \nu_{1,2} n+\psi_{1,2}$, and $\nu_{1,2}$ and $\psi_{1,2}$ are the tunes and 
initial phases of the normal modes. 
The initial phases $\psi_{1,2}$ are equal for all BPMs.
Typically in a storage ring the linear coupling is weak, in which case the observed $x$-motion is dominated 
by one normal mode and the $y$-motion by the other. 
For each transverse plane we call the dominant mode the primary mode and the other mode the secondary mode. 
The tunes of the primary modes are close to the uncoupled tunes for the corresponding planes. 
For the convenience of discussion, we refer the horizontal primary mode as normal mode 1 and 
the vertical primary mode as normal mode 2. 

The linear coupled motion of betatron coordinates $X=(x, x',y,y')^T$ at any location of the ring can be predicted with 
the one-turn transfer matrix ${\bf T}$. Diagonalizing the transfer matrix, one can relate 
betatron coordinates to normal mode coordinates 
 \begin{eqnarray}\label{eq:XPTheta}
{\bf \Theta }&=& \left( \begin{array}{c}
\sqrt{2J_1}\cos\Phi_1 \\
-\sqrt{2J_1}\sin\Phi_1 \\
\sqrt{2J_2}\cos\Phi_2 \\
-\sqrt{2J_2}\sin\Phi_2 
\end{array} \right) 
\end{eqnarray}
via a transformation ${\bf X}={\bf P}{\bf \Theta}$, 
where $J_{1,2}$ and $\Phi_{1,2}$ are the action and phase variables for the two normal modes, respectively~\cite{LuoYun2004}.  
In particular, the position coordinates $x$ and $y$ are given by
\begin{eqnarray}
x &=& p_{11}\sqrt{2J_1}\cos\Phi_1+\sqrt{2J_2}(p_{13}\cos\Phi_2-p_{14}\sin\Phi_2), \nonumber \\
y &=& \sqrt{2J_1}(p_{31}\cos\Phi_1-p_{32}\sin\Phi_1) + p_{33}\sqrt{2J_2}\cos\Phi_2, \nonumber \\ \label{eq:xyPhi}
\end{eqnarray}
where the $p_{ij}$ coefficients are elements of matrix ${\bf P}$ and by choice of the initial values of phase variables
$\Phi_{1,2}$, we have $p_{12}=p_{34}=0$~\cite{LuoYun2004}. 
Not considering damping of the coherent motion (e.g., due to decoherence), the action variables are constants of motion. 
The phase variables $\Phi_{1,2}$ advances from one location to another and the phase advances for a full turn are 
$2\pi\nu_{1,2}$. 

Clearly the measured beam motion in Eq.~(\ref{eq:ICAmodes}) and the model predicted motion in Eq.~(\ref{eq:xyPhi}) 
represent the same physical process and are separated in the same form. 
The amplitudes and phase advances of the two normal modes on the two transverse planes in the two equations should be equal. 
Equating the amplitudes, we obtain
\begin{eqnarray}\label{eq:Ampli1}
\sqrt{A^2+B^2} &=& \sqrt{2J_1}p_{11},  \\  \label{eq:Ampli2}
\sqrt{c^2+d^2} &=& \sqrt{2J_2}\sqrt{p_{13}^2+p_{14}^2},  \\ \label{eq:Ampli3}
\sqrt{C^2+D^2} &=& \sqrt{2J_2}p_{33}, \\ \label{eq:Ampli4}
\sqrt{a^2+b^2} &=& \sqrt{2J_1}\sqrt{p_{31}^2+p_{32}^2}.
\end{eqnarray}
The $J_{1,2}$ constants can be calculated by averaging the values derived from the amplitudes of 
the primary modes, i.e., using Eqs.~(\ref{eq:Ampli1}) and (\ref{eq:Ampli3}). 
Aside from constant initial phase, the phase advances can also be equated, leading to  
\begin{eqnarray}\label{eq:Phase1}
\tan^{-1}\frac{B}{A} &=& {\rm Mod}_{2\pi} (\Phi_1), \\ \label{eq:Phase2}
\tan^{-1}\frac{d}{c} &=& {\rm Mod}_{2\pi} (\Phi_2+\tan^{-1}\frac{p_{14}}{p_{13}}),  \\ \label{eq:Phase3}
\tan^{-1}\frac{b}{a} &=& {\rm Mod}_{2\pi} (\Phi_1+\tan^{-1}\frac{p_{32}}{p_{31}}),  \\ \label{eq:Phase4}
\tan^{-1}\frac{D}{C} &=& {\rm Mod}_{2\pi} (\Phi_2),  
\end{eqnarray}
where ${\rm Mod}_{2\pi}$ indicates taking modulus of $2\pi$ and we have made use of the 
fact that the value of arctangent can be uniquely determined within [0,$2\pi$) when both 
sine and cosine of an angle is known.

The phase advances of the normal modes $\Phi_{1,2}$ at the BPMs can be calculated with the lattice model. 
The ${\bf P}$ matrix can be computed from the one-turn transfer matrix at the BPM 
with the numeric procedure given in Ref.~\cite{LuoYun2004} or alternatively 
with equation ${\bf P} = {\bf VU}$, with matrices ${\bf V}$, ${\bf U}$ calculated with formulas given in Ref.~\cite{SaganRubin}. 

Eqs.~(\ref{eq:Ampli1})-(\ref{eq:Phase4}) apply at each BPM location. They relate quantities measured from 
TbT BPM data to quantities that can be calculated with the lattice model. 
In reality, the actual machine lattice deviates from the ideal model and there are instrumentation errors in 
the measurements. Therefore,  Eqs.~(\ref{eq:Ampli1})-(\ref{eq:Phase4}) are not exactly satisfied. 
However, one could adjust parameters in the lattice model and the diagnostics to minimize the differences 
between the measurements and the model prediction through a least-square problem. 
Based on these equations, an approach to calibrate the lattice model with TbT BPM data can be developed 
in a manner similar to the LOCO method, except here the input data used are different. 

Dispersion function measurements can be done quickly. They provide additional constraints to the lattice model 
and are themselves targets to be controlled through optics and coupling correction. 
Therefore, dispersion functions are also included in the least-square fitting problem. 

Instrumentation errors include BPM gain errors and BPM rolls. 
For each BPM, the measured beam position is related to actual beam position via a transformation
\begin{eqnarray}\label{eq:BPMgr}
\left( \begin{array}{c}
x \\ y
\end{array} \right)_{\rm meas} = {\bf B}
\left( \begin{array}{c}
x \\ y
\end{array} \right)_{\rm beam}, 
\end{eqnarray}
with 
\begin{eqnarray}
{\bf B} &=& 
\left( \begin{array}{cc}
\cos \theta & \sin \theta \\
-\sin \theta & \cos \theta
\end{array} \right)  
\left( \begin{array}{cc}
g_x & 0\\
0 & g_y 
\end{array} \right)
\end{eqnarray}
where $g_{x,y}$ and $\theta$ being the BPM gains and BPM roll, respectively. 
Transformation in Eq.~(\ref{eq:BPMgr}) is applied to both the transformation matrix ${\bf P}$ and the model dispersion functions 
before they are compared to the measurements. 
In LOCO a fourth parameter is included in matrix ${\bf B}$ to describe BPM transformation errors~\cite{SafranekLOCO}. 
We may do the same in the future in our method. 

The objective function to be minimized can be defined as 
\begin{eqnarray}\label{eqChi2}
f({\bf q}) = \chi^2 
 = \sum_{i,k} w_k \Big( \frac{d^{\rm meas}_{ik} - d^{\rm model}_{ik} }{\sigma_{ik}}\Big)^2,
\end{eqnarray}
where ${\bf q}$ is a vector that holds all fitting parameters, summation index $i$ stands for  
BPM $i$, $k$ stands for the $k$'th type of data,  
$d^{\rm meas}_{ik}$ and  $d^{\rm model}_{ik}$ are the measured and model values for data, respectively, 
${\sigma_{ik}}$ is the uncertainty level for data point $d^{\rm meas}_{ik}$, 
and $w_k$ is the weight factor for data type $k$. 
The fitting parameters include horizontal and vertical BPM gains, BPM rolls, quadrupole gradients, 
and skew quadrupole gradients. 
For each BPM, the fitting data include 
the amplitude of the horizontal primary mode ($k=1$), 
the amplitude of the vertical primary mode  ($k=2$),
the amplitude of the horizontal secondary mode  ($k=3$), 
the amplitude of the vertical secondary mode ($k=4$),  
horizontal dispersion  ($k=5$), vertical dispersion  ($k=6$), 
cosine  ($k=7$) and sine  ($k=8$) of the phase advance of the horizontal primary mode,
cosine  ($k=9$) and sine  ($k=10$) of the phase advance of the horizontal secondary mode,
cosine  ($k=11$) and sine  ($k=12$) of the phase advance of the horizontal primary mode,
and cosine  ($k=13$) and sine  ($k=14$) of the phase advance of the horizontal secondary mode. 
The betatron tunes can also be included in the objective function as fitting data. 
The weight factor may be used to adjust the relative weights of different types of data. 

TbT BPM data and orbit response matrix data  sample the linear optics and coupling of the storage ring 
in a similar manner, i.e., by perturbing the beam orbit in the transverse phase space area 
around the closed orbit. 
Orbit response matrix data are measured with closed orbit, which typically has higher precision than TbT orbit 
due to averaging over many turns. 
However, in TbT BPM data analysis, hundreds to thousands turns can be used, which statistically compensates 
the loss of precision for a single data point. 
In fact, because the advanced data analysis technique, ICA, is used to extract the betatron modes with 
data from all BPMs while filtering out BPM noise and contributions of other physical processes, 
it is possible that TbT BPM data have a statistical advantage over orbit response matrix 
data for optics and coupling measurements. 

It is known that quadrupole parameters in a storage ring lattice could have similar 
patterns in their perturbations to the optics and it may be difficult to separate their 
contributions to optics errors through fitting the LOCO data~\cite{HuangICFA44}. 
The same would be true to our TbT data based fitting method. 
Therefore, the constrained fitting technique proposed for LOCO in Ref.~\cite{HuangICFA44} 
should be used here, too. 

\section{Simulation \label{secSimul}} 
We have done simulation studies to demonstrate the applicability of the proposed method and 
the results are reported in Ref.~\cite{HuangYangIPAC15}. 
Lattice models for the SPEAR3 and NSLS-II storage rings are used in those studies. 
Quadrupole and skew quadrupole errors are planted into the lattice. 
TbT BPM data are generated by particle tracking, adding BPM gain and roll errors, 
and adding BPM noise. 
After applying ICA and fitting as described in the previous section, the BPM gain and roll errors 
are accurately recovered for both lattice models. 
The quadrupole and skew quadrupole errors are recovered for the SPEAR3 lattice. 
For the NSLS-II lattice, because of the degeneracy issue, only an approximately equivalent solution is 
obtained for the quadrupole and skew quadrupole errors. 
The optics error (beta beat) and coupling (ratio of projected emittance $\epsilon_y/\epsilon_x$) for the 
fitted lattice are very close to the target lattice, which indicate the fitted lattice is 
a good approximation to the latter. 

\section{Experiments \label{secExperi}} 
We applied our ICA based method to the NSLS-II storage ring to 
experimentally measure and correct the storage ring optics and coupling. 
NSLS-II storage ring is a new third generation storage ring light source that has been recently 
commissioned. A few selected parameters of the NSLS-II storage ring are shown in Table~\ref{tabParaNSLS2}. 
The storage ring is equipped with 180 BPMs, all of which are capable of taking turn-by-turn data~\cite{VetterNSLS2BPM}. 
\begin{table}[hbt] 
\caption{ Selected Parameters of NSLS-II }
\label{tabParaNSLS2}
 \begin{center}  
  \begin{tabular*}{0.4\textwidth}%
     {@{\extracolsep{\fill}}l|c|l}
  \hline
  Parameters & Value & Unit \\
  \hline  
  Energy  & $3$  & GeV    \\
  Circumference & 792.0 & m \\
  \# of periods & 30 & \\
  betatron tunes $\nu_{x,y}$ & 33.22, 16.26 & \\
  RF frequency $f_{\rm rf}$ & 499.68 & MHz \\
  Momentum compaction $\alpha_c$ & $0.36\times10^{-3}$ & \\
  \hline
  \end{tabular*}
  \end{center}
\end{table}  

\begin{figure}[phbt]
\psfrag{rmsA}{\small{$A_{\rm rms}$~(mm)}}
\psfrag{FFT amplitude}{\small{FFT amplitude}}
\psfrag{tune}{\small{tune}}
\psfrag{nus}{\small{$\nu_s$}}
\psfrag{nux}{\small{$\nu_x$}}
\psfrag{nuy}{\small{$\nu_y$}}
\psfrag{unkown}{\small{drift}}
  \centering
  \includegraphics[width=2.5in]{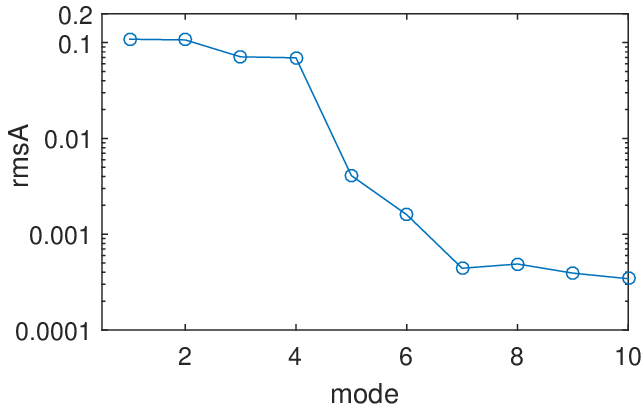}
    \vspace*{5mm}
  \includegraphics[width=2.5in]{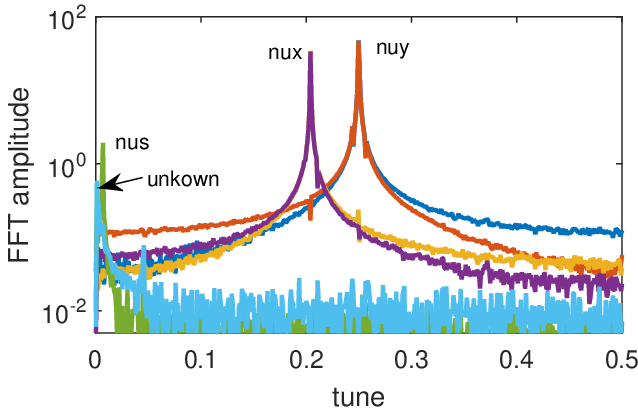}
    \caption{\label{figICAmodesIter1} Rms amplitude (top) and FFT spectrum (bottom) of the first six ICA modes from 
      the TbT BPM data of iteration 1.  The FFT spectrum is scaled by the rms amplitude for each mode.  }
\end{figure}

In the experiment, errors were initially inserted to the the machine lattice by adding random 
errors to magnet strengths. 
Coherent motion was excited simultaneously on the horizontal and vertical planes with the pingers 
before TbT BPM data were taken. 
The oscillation amplitudes for BPMs with large beta function values were about 0.2 mm on the horizontal plane 
and about 0.3 mm on the vertical plane, respectively. 
In addition, dispersion data were taken by shifting the rf frequency by 500 Hz. 
1024 turns of BPM data were used for the ICA analysis. 
Data from both planes form a $360\times1024$ matrix.  We kept 10 leading singular value modes 
for ICA analysis. 
The first four ICA modes are the betatron modes. The fifth mode is the synchrotron mode. 
The sixth mode shows a slow drift on its temporal pattern and its source is unknown. The other modes
are mostly within the noise level. 
FIG.~\ref{figICAmodesIter1} shows the amplitude (rms values for all BPMs) of all ten ICA modes
and the FFT spectrum of the first six modes.

\begin{figure}[phbt]
\psfrag{Dx}{\small{$D_x$~(m)}}
\psfrag{Dy}{\small{$D_y$~(m)}}
\psfrag{spos}{\small{$S$~(m)}}
  \centering
  \includegraphics[width=3.2in]{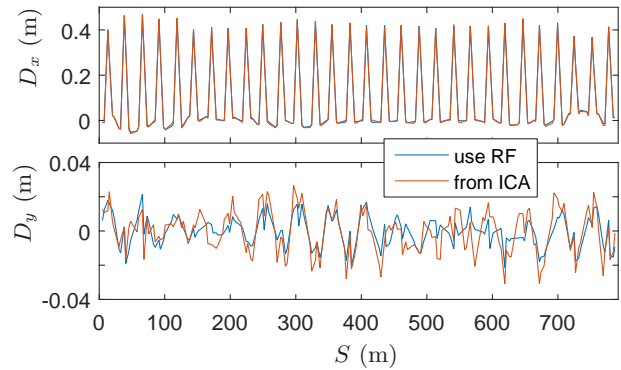}
    \caption{\label{figICADisp} Horizontal (top) and vertical (bottom) dispersion functions measured 
    with rf frequency shift (``use RF'') or from the TbT data through ICA (``from ICA'').  }
\end{figure}

The spatial pattern of the synchrotron mode (mode 5) is proportional 
to dispersion and can be used to measure 
the horizontal and vertical dispersion functions. 
The energy oscillation amplitude is determined by requiring the horizontal rms dispersion to be equal to 
the model value, from which the energy oscillation amplitude was found to be $\delta_m=0.3\times10^{-4}$. 
The measured synchrotron tune is $\nu_s=0.00657$. 
Dispersion functions measured from the TbT data are compared to results from the usual closed orbit based method 
in FIG.~\ref{figICADisp}. 
Reasonable agreement is seen for both horizontal and vertical dispersions. 

\begin{figure}[pbt]
\psfrag{spos}{\small{$S$~(m)}}
\psfrag{delphix}{\small{$\Delta\phi_x$}}
\psfrag{delphiy}{\small{$\Delta\phi_y$}}
  \centering
  \includegraphics[width=3.2in]{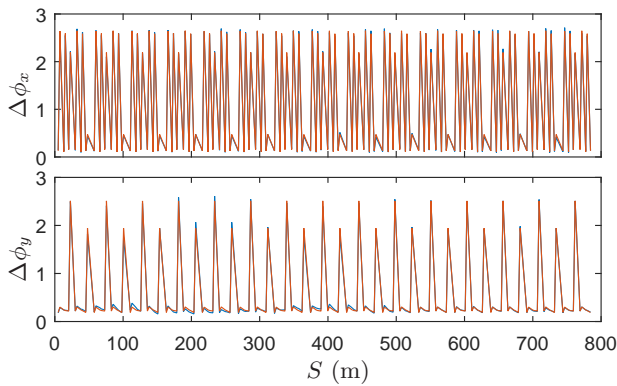}
    \caption{\label{figICAdPhaseIter1} Phase advances between adjacent BPMs from ICA measurements before corrections (blue) 
				are compared to the ideal model (red) for horizontal (top) and vertical (bottom) planes.  }
\end{figure}

\begin{figure}[pbt]
\psfrag{spos}{\small{$S$~(m)}}
\psfrag{diffxy}{\small{$\Delta \phi$ diff. (rad)}}
\psfrag{ddphix}{\small{hori }}
\psfrag{ddphiy}{\small{vert}}

  \centering
  \includegraphics[width=3.2in]{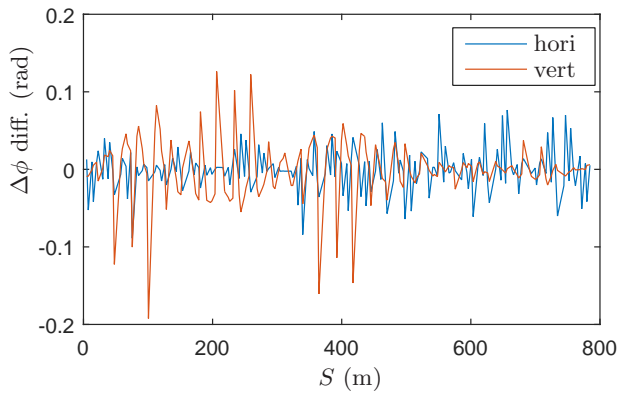}
    \caption{\label{figICADiffPhaseIter1} Differences between measured and model phase advances between adjacent BPMs 
    before corrections.  }
\end{figure}

From the spatial patterns of the ICA betatron modes, the betatron amplitudes and phase advances at the BPMs 
can be calculated~\cite{XHuangICA2005}. 
By requiring the average beta functions as derived from the primary modes at the BPMs to be equal to 
the values of the ideal model, we obtained the action variables of the coherent motion for the 
horizontal and vertical planes, which are $J_1=1.41\times10^{-9}$~m and   $J_2=2.82\times10^{-9}$~m, respectively. 
The fractional betatron tunes obtained from the temporal patterns are $\nu_x=0.2048$ and $\nu_y=0.2513$, 
respectively. 
The measured phase advances between adjacent BPMs are compared to the model values in FIG.~\ref{figICAdPhaseIter1} 
and their differences are shown in FIG.~\ref{figICADiffPhaseIter1}. 


\begin{figure}[pbt]
\psfrag{spos}{\small{$S$~(m)}}
\psfrag{ampli. ratio}{\small{amplitude ratio}}
  \centering
  \includegraphics[width=3.2in]{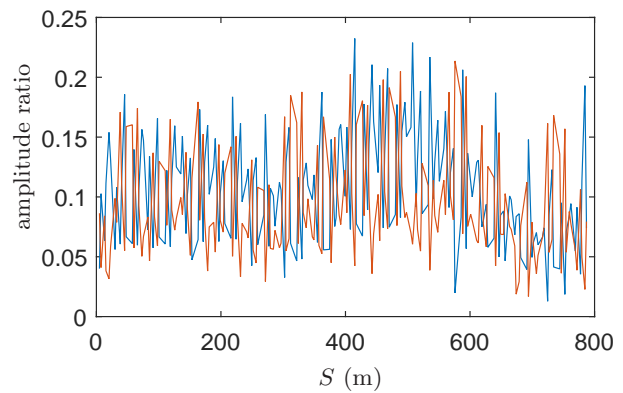}
    \caption{\label{figICAAmpRatioIter1} Amplitude ratio of the projected motion onto the two transverse planes for 
    the two normal modes, as defined in Eq.~(\ref{eqCouplingAmpRatio}), for TbT data before corrections.  }
\end{figure}

The ratio of the amplitudes of the projected motion of a normal mode onto the two transverse planes is a measure 
of the local coupling. From Eqs.~(\ref{eq:Ampli1}-\ref{eq:Ampli4}), we can define
\begin{eqnarray}
r_1 &=& \frac{\sqrt{a^2+b^2} }{ \sqrt{A^2+B^2}} =\frac{\sqrt{p_{31}^2+p_{32}^2}}{p_{11}}, \nonumber \\
r_2 &=& \frac{\sqrt{c^2+d^2} }{ \sqrt{C^2+D^2}} =\frac{\sqrt{p_{13}^2+p_{14}^2}}{p_{33}}.  \label{eqCouplingAmpRatio}
\end{eqnarray}
The amplitude ratio for the two normal modes at all BPMs are shown in FIG.~\ref{figICAAmpRatioIter1}. 

The ICA results are used for lattice model fitting as described in section~\ref{secMethod}. 
The fitting parameters are horizontal gains, vertical gains, and rolls for all 180 BPMs, 150 quadrupole parameters, 
and 30 skew quadrupole parameters. 
The quadrupole and skew quadrupole parameters are the same as the LOCO setup for NSLS-II~\cite{YangLOCOIPAC15}. 
The normalizing sigmas in Eq.~(\ref{eqChi2}) are set to 0.05~m$^{-1/2}$ for the betatron amplitudes, 0.005~m for 
dispersion functions, 0.005~rad for phase advances of the primary modes, and 0.02~rad for phase advances of the 
secondary modes. 
Constraints are added to all fitting parameters with the cost function proportional to the 2-norm of the 
corresponding column of the Jacobian matrix of each fitting parameter. 
$\chi^2$ normalized by the degree of freedom dropped from 140.9 to 1.8 in two fitting iterations 
for the first TbT data set. 

\begin{figure}[pbt]
\psfrag{spos}{\small{$S$~(m)}}
\psfrag{dBBx}{\small{$\Delta \beta_x/\beta_x$}}
\psfrag{dBBy}{\small{$\Delta \beta_y/\beta_y$}}
  \centering
  \includegraphics[width=3.2in]{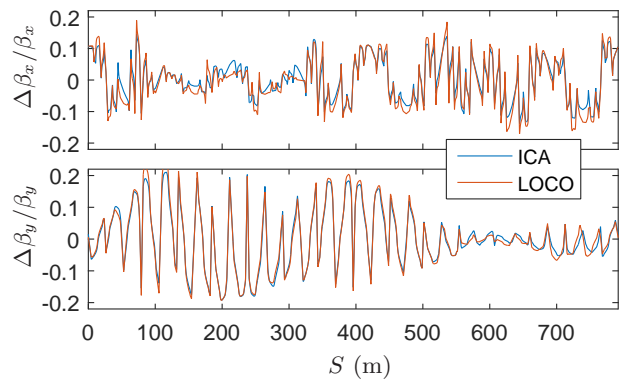}
    \caption{\label{figCmpdBBIter1} Comparison of beta beating of lattices fitted by TbT data (``ICA'') 
    and LOCO for horizontal (top) and vertical (bottom) beta functions before corrections. }
\end{figure}

\begin{figure}[pbt]
\psfrag{spos}{\small{$S$~(m)}}
\psfrag{epsratio}{\small{$\epsilon_y/\epsilon_x$}}
  \centering
  \includegraphics[width=3.2in]{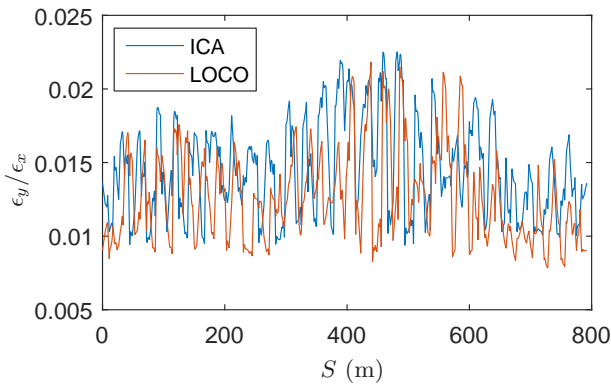}
    \caption{\label{figCmpEpsYIter1} Comparison of ratios of projected vertical emittance over horizontal 
    emittance     for lattices fitted by TbT data (``ICA'')     and LOCO before corrections. }
\end{figure}

Orbit response matrix data were also taken during the experiment under the same machine condition. 
The data were fitted for 
three iterations. The results are compared to the ICA fitting results. 
The optics of the fitted lattices are compared in FIG.~\ref{figCmpdBBIter1}. There was excellent agreement 
between ICA and LOCO in the distribution of beta beating, which clearly verifies that the two fitted lattices are 
nearly equivalent. 
FIG.~\ref{figCmpEpsYIter1} shows the distribution of the projected vertical emittance (normalized by the 
horizontal emittance) for the two fitted latices. Although not all details agree, the average emittance ratios 
are nearly the same. 


\begin{figure}[pbt]
\psfrag{dKK}{\small{$\Delta K/K$}}
\psfrag{DeltaK}{\small{$\Delta K$~(m$^{-2}$)}}
  \centering
  \includegraphics[width=3.0in]{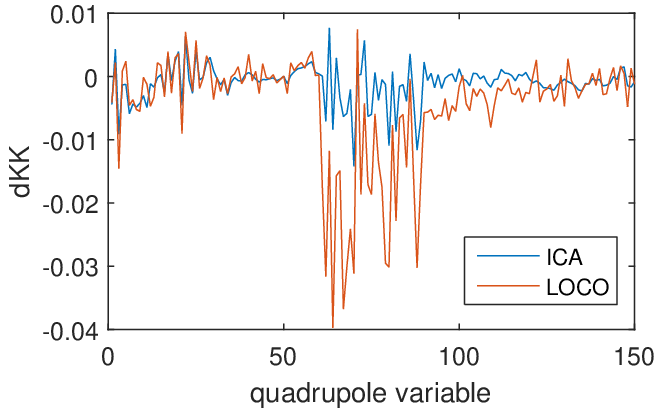}\par
  \vspace{4mm}
  \includegraphics[width=3.0in]{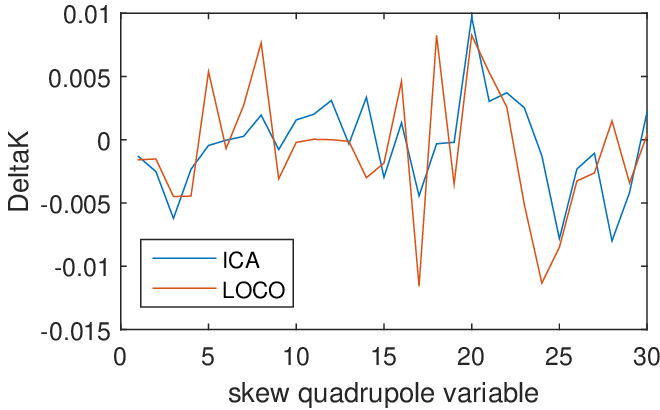}
    \caption{\label{figCmpdKKskewKIter1} Comparison of fitted quadrupole variables (top) and 
    skew quadrupole variables (bottom)  for lattices fitted by TbT data (``ICA'')     and LOCO for iteration 1. }
\end{figure}

The fitted lattice parameters for ICA fitting and LOCO are compared in FIG.~\ref{figCmpdKKskewKIter1}. 
Good agreement was seen for quadrupole variables 1-60 (QL1 and QL2) and 121-150 (QM2). 
But the fitted errors of quadrupole variables 61-120 (QH1 and QH2) for LOCO are larger. 
Since the fitted optics of the two methods are equivalent (see FIG.\ref{figCmpdBBIter1}), 
the differences in the fitted quadrupole errors indicate the cost functions for these variables in the 
LOCO setup need to be increased.  

\begin{figure}[hpbt]
\psfrag{iteration}{\small{Iteration}}
\psfrag{initial chi2}{\small{initial $\chi^2/N$}}
  \centering
  \includegraphics[width=3.0in]{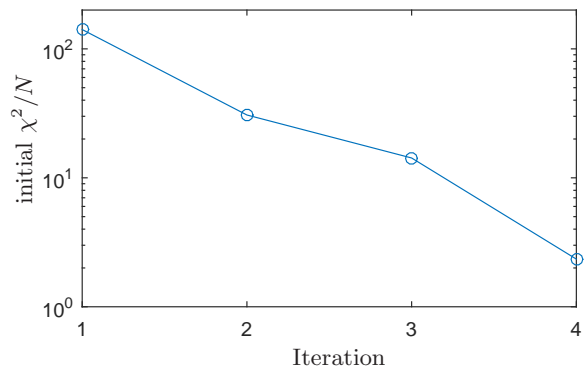}
    \caption{\label{figInitChi2} Initial normalized $\chi^2$ (before fitting) for the four TbT 
    BPM data sets taken during the experiment.  
     }
\end{figure}

The ICA fitting results of quadrupole and skew quadrupole variables were applied to the machine to 
correct optics and coupling. 
Then a new TbT BPM data set was taken, fitted, and used for correction. 
A total of three corrections were applied. 
A final TbT BPM data set and an orbit response matrix 
data set were taken after the corrections. 
FIG.~\ref{figInitChi2} shows the initial $\chi^2$ (before fitting) normalized by the degree of 
freedom for the four TbT BPM data sets. 
The reduction of initial $\chi^2$ indicates that the optics and coupling were being improved through the 
iterations of corrections. 

\begin{figure}[pbt]
\psfrag{spos}{\small{$S$~(m)}}
\psfrag{diffxy}{\small{$\Delta \phi$ diff. (rad)}}
\psfrag{ddphix}{\small{hori }}
\psfrag{ddphiy}{\small{vert}}
  \centering
  \includegraphics[width=3.2in]{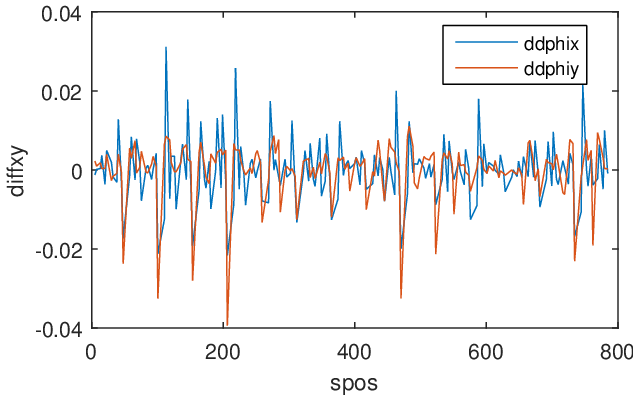}
    \caption{\label{figICADiffPhaseIter4} Differences between measured and model phase advances between adjacent BPMs 
    after corrections.  }
\end{figure}

\begin{figure}[pbt]
\psfrag{spos}{\small{$S$~(m)}}
\psfrag{Dx}{\small{$D_x$~(m)}}
\psfrag{Dy}{\small{$\phantom{a}D_y$}}
\psfrag{dDxDy}{\small{$D_y$, $\Delta D_x$~(m)}}
\psfrag{DeltaDx}{\small{$\Delta D_x$}}
  \centering
  \includegraphics[width=3.2in]{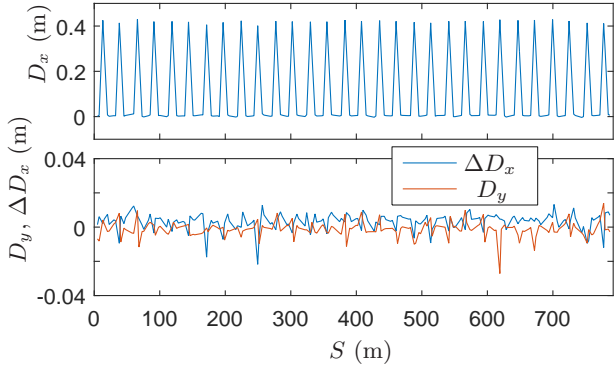}
    \caption{\label{figCmpDisperIter4} Measured horizontal  (top) and vertical (bottom,$ D_y$) dispersions 
    and the difference of horizontal dispersion with the model dispersion (bottom, $\Delta D_x$) after corrections. }
\end{figure}

A series of measures directly derived from the TbT BPM data or dispersion measurements before 
and after the ICA fitting based corrections are shown in Table~\ref{tabMeasDirect}. 
After corrections, the measured fractional betatron tunes are much closer to their nominal values. 
The rms values of the phase advance differences between measurements and models 
(shown in FIG.~\ref{figICADiffPhaseIter4}) are a factor of 3.6 and 5.3 smaller 
from the initial values (before corrections) for 
the two transverse planes, respectively. 
The measured dispersion functions after corrections are shown in FIG.~\ref{figCmpDisperIter4}. 
Compared to measurements before corrections (in FIG.~\ref{figICADisp}), 
significant improvements were made on both horizontal and vertical dispersions. 
The residual vertical dispersion includes sharp peaks that 
are due to rolls of BPMs at large horizontal dispersion locations.  

\begin{figure}[pbt]
\psfrag{spos}{\small{$S$~(m)}}
\psfrag{ampli. ratio}{\small{amplitude ratio}}
  \centering
  \includegraphics[width=3.2in]{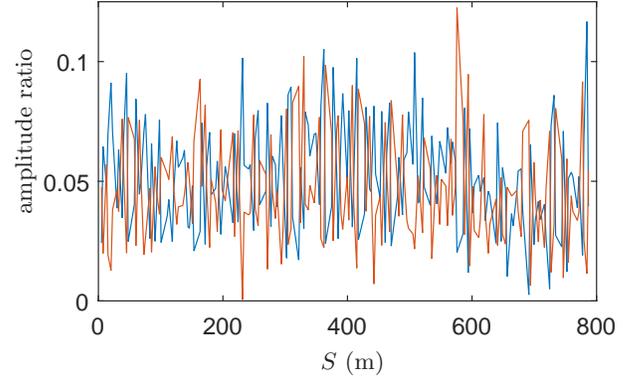}
    \caption{\label{figICAAmpRatioIter4} Amplitude ratio of the projected motion onto the two transverse planes for 
    the two normal modes for TbT data after corrections.  }
\end{figure}

The coupling amplitude ratios as defined in Eq.(\ref{eqCouplingAmpRatio}) after correction 
are shown in FIG.~\ref{figICAAmpRatioIter4}. 
The average ratios are reduced by about a factor of 2. 
The direct measures in Table~\ref{tabMeasDirect} 
clearly show that both the linear optics and linear coupling were 
significantly improved with the ICA fitting based corrections. 

\begin{figure}[pbt]
\psfrag{spos}{\small{$S$~(m)}}
\psfrag{dBBx}{\small{$\Delta \beta_x/\beta_x$}}
\psfrag{dBBy}{\small{$\Delta \beta_y/\beta_y$}}
  \centering
  \includegraphics[width=3.2in]{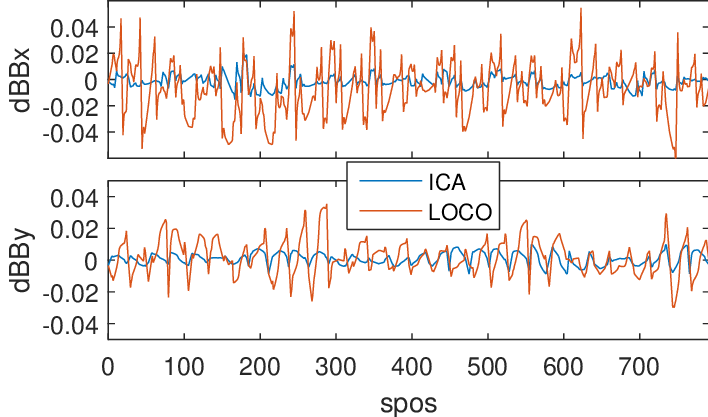}
    \caption{\label{figCmpdBBIter4} Comparison of beta beating of lattices fitted by TbT data (``ICA'') 
    and LOCO for horizontal (top) and vertical (bottom) beta functions after corrections. }
\end{figure}

\begin{figure}[pbt]
\psfrag{spos}{\small{$S$~(m)}}
\psfrag{epsratio}{\small{$\epsilon_y/\epsilon_x$}}
  \centering
  \includegraphics[width=3.2in]{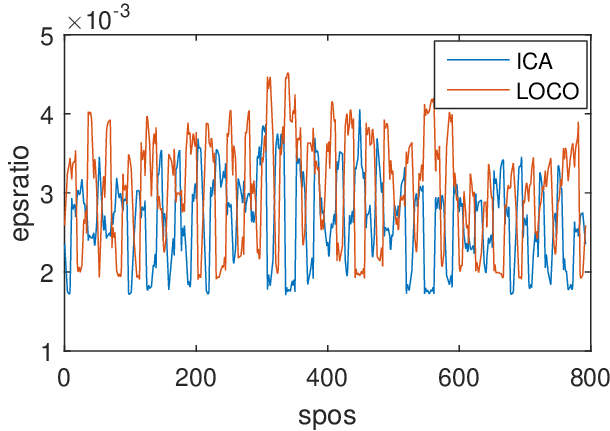}
    \caption{\label{figCmpEpsYIter4} Comparison of ratios of projected vertical emittance over horizontal 
    emittance     for lattices fitted by TbT data (``ICA'')     and LOCO after corrections. }
\end{figure}

The ICA fitting results from the final TbT BPM data are also compared to the fitting results of 
the final LOCO data. 
The rms beta beating and average coupling ratio obtained with the two methods before and 
after corrections are shown in Table~\ref{tabMeasFitting}. 
FIG.~\ref{figCmpdBBIter4} shows  beta beating  from the two methods after corrections. 
Rms beta beating from LOCO is bigger than ICA fitted values. 
The discrepancy between the two methods needs further investigation. 
Using the LOCO values, beta beating were reduced by a factor of 4.0 (horizontal) 
and 9.0 (vertical), respectively. 
Coupling ratios of the fitted lattices are compared in FIG.~\ref{figCmpEpsYIter4}. 
The reduction of coupling ratio is a factor of 5.5 according to ICA fitting and 
4.2 according to LOCO. 

\begin{table}[hpbt] 
\caption{ Direct lattice characterization before and after corrections. 
 }
\label{tabMeasDirect}
 \begin{center}  
  \begin{tabular*}{0.4\textwidth}%
     {@{\extracolsep{\fill}}l|cc}
  \hline
  Parameters & before & after \\
  \hline  
  fractional $\nu_x$  & 0.2048 & 0.2193  \\
  fractional $\nu_y$  & 0.2513 & 0.2602  \\
  rms $\Delta \phi_x$ diff. (rad)  & 0.0280 & 0.0077  \\
  rms $\Delta \phi_y$ diff. (rad)  & 0.0390 & 0.0074  \\
  rms $\Delta D_x$ (m) & 0.0171 & 0.0049  \\
  rms $D_y$ (m) & 0.0089 & 0.0046  \\
  mean $r_1$ & 0.106    & 0.052 \\
  mean $r_2$ & 0.094    & 0.046 \\
  \hline
  \end{tabular*}
  \end{center}
\end{table}  

\begin{table}[hpbt] 
\caption{ Fitted lattice parameters by ICA and LOCO before and after corrections.  
 }
\label{tabMeasFitting}
 \begin{center}  
  \begin{tabular*}{0.4\textwidth}%
     {@{\extracolsep{\fill}}l|cc|cc}
  \hline
   & before & & after &\\
  \hline
  			Parameters  & ICA & LOCO & ICA & LOCO\\
  \hline  
 rms $\Delta \beta_x/\beta_x$ & 0.0678  & 0.0780 & 0.0051 & 0.0194 \\
 rms $\Delta \beta_y/\beta_y$ & 0.0937 & 0.0991 & 0.0038 & 0.0110 \\
 mean $\epsilon_y/\epsilon_x$ &  0.0147 &  0.0129 & 0.0027  & 0.0031 \\
  \hline
  \end{tabular*}
  \end{center}
\end{table}  

In FIG.~\ref{figCmpEpsYIter4} we show a comparison of fitted BPM gains for the two methods. 
There are discrepancies at some locations. 
But at more locations the two methods agree.



\begin{figure}[pbt]
\psfrag{BPM}{\small{BPM}}
\psfrag{gx}{\small{$g_x$}}
\psfrag{gy}{\small{$g_y$}}
  \centering
  \includegraphics[width=3.2in]{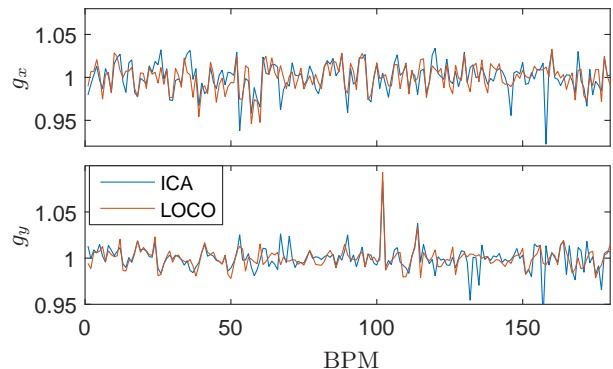}
    \caption{\label{figCmpBgainsIter4} Comparison of fitted BPM gains by TbT data (``ICA'') 
    and LOCO for horizontal (top) and vertical (bottom) BPMs after corrections. }
\end{figure}

\section{\label{secConclu}Conclusion}
We propose a method to use turn-by-turn (TbT) BPM data to simultaneously correct 
linear optics and linear coupling for circular accelerators. 
The independent component analysis (ICA) is applied to extract the betatron normal modes 
and their projection on the two transverse planes~\cite{XHuangICA2005}. 
The betatron amplitudes and phase advances of the projected modes are fitted to the 
lattice model. Linear optics and coupling of the machine can be calculated with the fitted model. 
Correction is made by reversing the fitted errors of the quadrupole and skew quadrupole variables. 

This new method has been successfully demonstrated on the NSLS-II storage ring. 
In the experiment we reduced rms beta beat from 7.8\% 
(horizontal) and 9.9\%
(vertical) to 1.9\% 
(horizontal) and 1.1\% 
for the two transverse planes, respectively, using LOCO results as references. 
The coupling ratio was reduced from 1.3\%
to 0.3\%. 

Because TbT can be taken within seconds and with little perturbation to the 
stored beam (in our experiment coherent oscillation amplitude is below 0.3~mm), 
this method has a great advantage over the orbit 
response matrix based method, LOCO~\cite{SafranekLOCO}.
For storage ring machines, this method may be used to check and correct optics 
during operation using data taken with injection transients. 
It may also improve efficiency for machine startup and machine studies 
as less time is needed for lattice correction.

\begin{acknowledgments}
We thank Yongjun Li for preparing the initial condition of the 
machine for the experiment. 
We thank James Safranek for reading the manuscript. 
Author X. Yang thanks Victor Smalyuk for many useful discussions. 
The study is supported by U.S. DOE under Contract No. DE-AC02-98CH10886 (BNL) and 
Contract No. DE-AC02-76SF00515 (SLAC). 
\end{acknowledgments}

\bibliography{da_ref}

\end{document}